\documentclass[sigconf]{acmart}

\AtBeginDocument{%
  \providecommand\BibTeX{{%
    \normalfont B\kern-0.5em{\scshape i\kern-0.25em b}\kern-0.8em\TeX}}}
    
\setcopyright{acmcopyright}
\copyrightyear{2018}
\acmYear{2018}
\acmDOI{10.1145/1122445.1122456}

\acmConference[ICSE 2020 SEIP]{ICSE 2020 SEIP: Software Engineering in Practice}{2020}{Seoul, Korea}

\usepackage{algorithm}
\usepackage[noend]{algpseudocode}
\usepackage{amsmath}

\newcommand{\etal}{\textit{et al}. }

\newcommand{\eg}{\textit{e}.\textit{g}. }
\newcommand{\ignore}[1]{}
\newcommand{\MID}{SIG}

\begin{document}

\title{Debugging Crashes using \textit{Continuous} Contrast Set Mining}

\author{Rebecca Qian}
\authornotemark[1]
\affiliation{
  \institution{Facebook, Inc.}
  \country{U.S.A.}
}
\email{rebeccaqian@fb.com}

\author{Yang Yu}
\affiliation{
  \institution{Purdue University}
  \country{U.S.A.}}
\email{yu577@purdue.edu}

\author{Wonhee Park}
\affiliation{
  \institution{Facebook, Inc.}
  \country{U.S.A.}
}
\email{wonheepark@fb.com}

\author{Vijayaraghavan Murali}
\affiliation{
 \institution{Facebook, Inc.}
 \country{U.S.A.}}
\email{vijaymurali@fb.com}

\author{Stephen Fink}
\affiliation{
  \institution{Facebook, Inc.}
  \country{U.S.A.}}
\email{stephenfink@fb.com}

\author{Satish Chandra}
\affiliation{\institution{Facebook, Inc.}
\country{U.S.A.}}
\email{schandra@acm.org}

\renewcommand{\shortauthors}{Qian, et al.}

\begin{abstract}
  Facebook operates a family of services used by over two billion people daily on a huge
  variety of mobile devices.  Many devices are configured to upload crash reports should the app crash for any reason.   Engineers monitor and triage
  millions of crash reports logged each day to check for bugs, regressions, and
  any other quality problems.   Debugging groups of crashes is a manually intensive process that requires deep domain expertise and close inspection of traces and code, often under time constraints. 
   
   We use contrast set mining, a form of discriminative pattern mining, to learn what distinguishes one group of crashes from another. Prior works focus on discretization to apply contrast mining to continuous data. We propose the first direct application of contrast learning to continuous data, without the need for discretization. We also define a weighted anomaly score that unifies continuous and categorical contrast sets while mitigating bias, as well as uncertainty measures that communicate confidence to developers. We demonstrate the value of our novel statistical improvements by applying it on a challenging dataset from
   Facebook production logs, where we achieve \textbf{40$\mathbf{x}$ speedup} over baseline approaches using discretization.

\end{abstract}

\ccsdesc[500]{Software and its engineering~Software reliability}

\keywords{crash analysis, descriptive rules, rule learning, contrast set mining, emerging patterns, subgroup discovery, multiple hypothesis testing}

\maketitle

\section{Introduction}
\label{sec:intro}
In commercial software development, despite significant
investment in software quality processes including static and dynamic analysis, code reviews and testing, defects still slip through and cause crashes in the field.
Fixing these crashes remains a manually intensive process, demanding deep domain expertise and detailed analysis of traces and code.

Large software organizations often deploy automated crash triage systems, which capture error logs when a mobile client crashes.
These logs contain hundreds of key-value pairs with metadata about the app's execution environment, such as the mobile OS version or app build, and possibly a trace of where a crash occurred.
Once captured, an automated system usually groups crash logs into categories (e.g., by a hash on a descriptive value, or through more sophisticated clustering) and then assigns each category to on-call developers. 
If categorization were perfect, each crash in a category would arise from the same root cause.

Often, developers trying to resolve a group of crashes want to know what distinguishes a particular group of crashes.
For instance, developers ask: ``does this group of crashes occur disproportionately in build version X?'', or ``does this group of crashes occur disproportionately for users from country Y?''.
One simple way to describe a group of crashes is to use standard statistical tests regarding the distribution of features among members of the group.
For example, one could test if \texttt{country:Y} appears statistically more frequently in one group of crashes than in the whole population.

One limitation of standard statistical tests is that they may not reveal patterns involving interactions among multiple
features. 
Our crash data includes many dimensions of features, with a mix of categorical, discrete, and continuous values.   
Additionally, we need to generate {\em interpretable} insights that a human can comprehend, which rules out standard dimensionality reduction techniques (\eg principal component analysis) which tend to compute complex, unintuitive factors.

Recently, Castelluccio~\etal~\cite{Castelluccio:2017} proposed using {\em contrast set mining} to extract insights from multi-dimensional data about crashes.
Contrast set mining (CSM)~\cite{Bay:1999} is a form of discriminative pattern mining that attempts to discover significant patterns that occur with disproportionate frequencies in different groups.
It explores the space of feature sets, i.e., sets of conjunctive feature-value pairs, looking for deviations from expected distributions.
For example, the feature set $\{{\sf build\_version: X, country: Y}\}$ is interpreted as the value of the build version feature being X and country being Y.
CSM models the expected distributions of these sets from the general population of data points, i.e., independent of any particular group.
Then, given a particular group of points, if the distribution of a feature set in that group differs significantly from its expected value, it is labeled as a {\em contrast set}.
The notion of differing significantly is defined by an explicit statistical test, and denotes a degree to which the contrast set is anomalous.
Castelluccio~\etal's application of CSM produced relevant hints to developers regarding groups of user-reported bugs, helping them fix bugs faster.
It also helped uncover systematic breakages by detecting anomalous attribute-value pairs.

Thus far, most applications of CSM consider only categorical data.
When applying CSM to discrete and continuous variables, researchers typically \textit{discretize} data by sorting values into buckets, resulting in a limited number of possible states.
For example, $\{{\sf process\_uptime: (0, 2000)}\}$ represents a discretized feature that denotes instances where a process ran for less than $2000$ milliseconds.
Discretization has notable drawbacks; it does not scale well to large datasets with thousands of features as it leads to an explosion in the number of feature-value pairs.
Moreover, if the feature ranges are strongly skewed, the discretized bins do not capture the distribution well.
Discretized ranges are also often unintuitive, causing resulting contrast sets to be difficult to interpret. 

To address these drawbacks, we propose several improvements to CSM to extend it to continuous features and other mixed data types, such as event sequences, {\em without discretization}. 

We demonstrate its effectiveness by applying it to a class of hard bugs: app deaths from iOS out-of-memory crashes (OOMs).
These OOMs are hard to resolve since they do not provide logs with stack traces.
Instead, they are annotated with {\em user navigation logs}, i.e., sequences of events that a user navigated, ordered chronologically, before the crash occurred.
As we will describe later, we compute a continuous vector-space encoding of these navigation logs as a technique to enable tractable analysis of this high dimensional data, while still enabling accurate localized information about a potential root cause.

Towards these goals, our key contributions are the following:

\begin{itemize}

\item We propose Continuous Contrast Set Mining (CCSM), the first direct application of contrast mining to continuous data without discretization.
\item We evaluate its effectiveness on a dataset containing $60k$ iOS OOM crashes: using CCSM, on average we generated 1120 contrast sets in 458 seconds.
This is a \textbf{40$\mathbf{x}$ speedup} over a naive discretization approach. We also evaluate the usefulness of CCSM towards debugging software issues in an industrial setting.
\item We provide a formal definition of contrast set quality, allowing us to rank contrast sets:
\begin{itemize}
    \item A weighted anomaly score for categorical data that can be viewed as a transformed Euclidean distance measure between observed and expected values, and
    \item A normalized anomaly score that unifies continuous and categorical contrast sets while mitigating bias. 
\end{itemize}
\item We propose uncertainty measures based on confidence intervals evaluating effect size and difference in means to provide signals to developers on actionability.

\end{itemize}

The rest of the paper is organized as follows.
Section~\ref{sec:overview} gives an overview of the previously proposed algorithm for CSM and its limitations when working with continuous features.
Section~\ref{sec:algorithms} describes our proposed algorithm, CCSM, that handles continuous features and a definition of anomaly score that unifies categorical and continuous contrast sets.
Section~\ref{sec:csm_industry} discusses how contrast sets from CCSM can aid software debugging in practice in the industry.
Section~\ref{sec:results} presents experimental results evaluating our algorithm, including preliminary experience with it.
Section~\ref{sec:discussion} describes threats to the validity of our study.
Finally, Sections~\ref{sec:related} and~\ref{sec:conclusion} discuss related works and future directions, respectively.

\section{Overview}
\label{sec:overview}

In this section, we describe the CSM problem and summarize the existing algorithm for CSM on categorical variables, namely STUCCO~\cite{Bay:1999}.
We then discuss what kind of continuous features arise in crashes and limitations of using discretization to apply STUCCO to the continuous domain.

\subsection{STUCCO Contrast Set Mining Algorithm}

The objective of CSM is to find statistically meaningful differences between groups. In CSM, we start with a categorical dataset partitioned into mutually exclusive groups. A candidate contrast set is a conjunction of attribute-value pairs, where an attribute is a database field and the value is one of a range of values that field can take on, \eg \texttt{country=IN}.

For a contrast set $X$ and group $G$, the support $S(X, G)$ is the \textit{percentage of vectors in group $G$ for which the contrast set $X$ is true}. We want to find "interesting" contrast sets whose support differs \textit{meaningfully} across groups. For differences in support to be meaningful, the contrast set must be both \emph{significant} and \emph{large}. More formally, contrast sets must satisfy two conditions,

\begin{equation}
    \label{significant}
    \exists ij \text{ s.t. } P(X | G_i) \neq P(X | G_j)
\end{equation}
and
\begin{equation}
    \label{large}
    max_{ij} |S(X, G_i) - S(X, G_j)| \geq \delta
\end{equation}
where $P(X|G_i)$ is the likelihood of observing set $X$ for group $G_i$ and $\delta$ is the user defined \textit{minimum support difference}. Equation ~\eqref{significant} tests a contrast set for statistical significance, and Equation ~\eqref{large} checks for largeness, i.e., that the support of the contrast set differs by a certain threshold for at least two groups.

\begin{figure}[h]
  \centering
  \includegraphics[width=\linewidth]{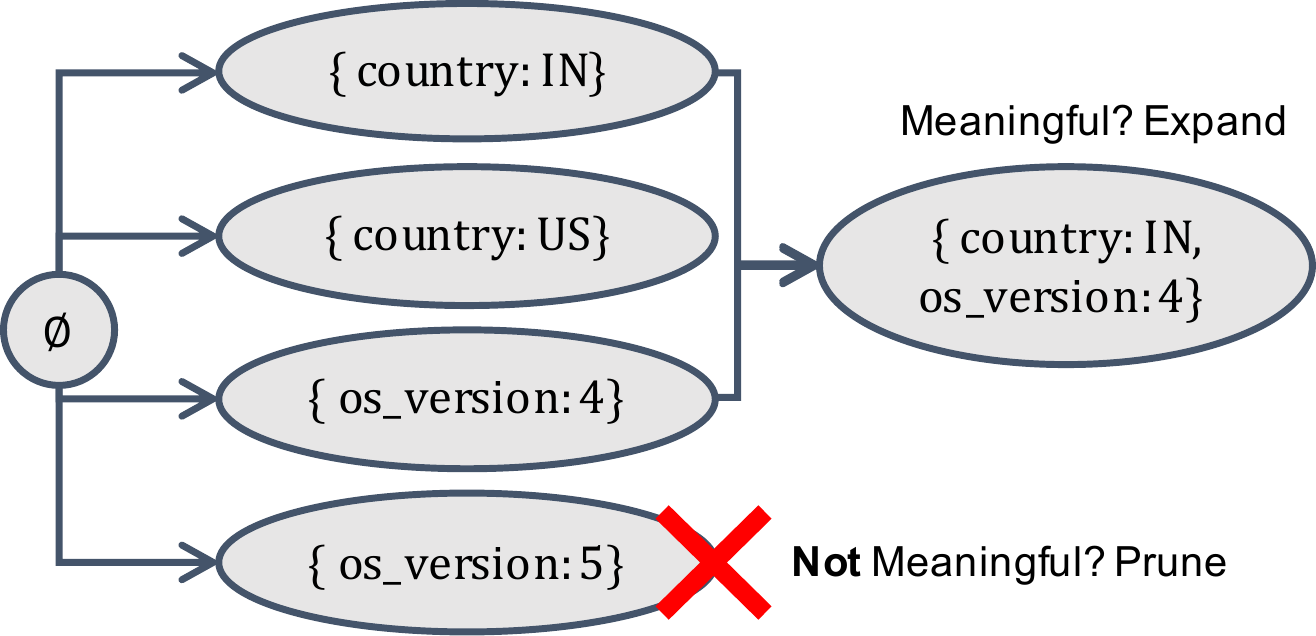}
  \caption{Sample generation of initial candidate sets}
  \label{fig:candidates}
\end{figure}

For the baseline, we implemented the STUCCO algorithm, which casts contrast set mining as a tree search problem~\cite{Bay:1999}. Starting with an empty root node, we begin by enumerating all attribute-value pairs in the dataset. Figure ~\ref{fig:candidates} shows the initial candidates for a toy dataset with two columns ($country$, $os\_version$) that each can take on two values.

For each candidate node, we scan the dataset and count support for each group. We then examine whether the node is significant and large. In STUCCO, a contrast set is statistically significant if it passes a two-way Chi Squared test. The \textit{null hypothesis} is that the support of a contrast set is equal across all groups, i.e., independent of group membership. The Chi Squared test evaluates this hypothesis by analyzing expected and observed frequencies, taking into account factors such as the number of observations, group sizes and variance.

Nodes that fail either condition ~\eqref{significant} or ~\eqref{large} are pruned. After testing all candidates in a given level, we generate children from the surviving nodes by forming conjunctions of attributes. We use a canonical ordering of attributes to avoid redundant combinations. For example, assuming no nodes are pruned, the children of contrast set $\{{\sf country: IN}\}$  are $\{{\sf country: IN, os\_version: 4.0}\}$  and $\{{\sf country: IN, os\_version: 5.0}\}$.

The generated child nodes become the new candidate set. In addition to significant and large conditions, we implement a variety of pruning techniques as described by Castelluccio \etal~\cite{Castelluccio:2017}. We repeat the above process of testing and generating contrast sets, until no new child nodes can be formed. As in~\cite{Castelluccio:2017}, we reduce the likelihood of type 1 errors (false positives) in statistical testing by applying the Bonferroni correction, which lowers critical values as the tree depth increases.

Contrast set mining has several advantages when compared to other techniques used to mine feature sets, such as Decision Trees. Broadly speaking, the main advantage of decision trees is their flexibility; in comparison, a single Chi Squared test only tests one hypothesis at a time. STUCCO does multiple hypothesis testing efficiently by casting it as a tree search problem, allowing production usage at scale. Compared to CSM, limitations of Decision Tree algorithms include the lack of statistical significance testing, which does not provide guarantees on split quality. Additionally, as with other greedy algorithms, the order of decisions impacts the results.

Next, we describe the setting in which we wish to apply CSM, and the limitations of the standard STUCCO algorithm in this setting.

\subsection{Continuous Features of Crash Reports}
With billions of active mobile users, Facebook must monitor and maintain the health of mobile apps
at huge scale.

When a crash occurs, a snapshot of the mobile client and app level information is logged into a crash report, which is then uploaded to a server.
A crash report often includes the stack trace associated with the crash, which is one of the most important signals for a developer debugging the crash.

For certain classes of crashes, however, the stack trace is unavailable or difficult to obtain.
For instance, when an out-of-memory error (OOM) occurs in unmanaged code, the OS kills the app 
and does not have memory to snapshot a stack trace.
In other crashes from native code, the stack trace may not contain debugging symbols and conveys
little interpretable information.
For these classes of crashes, termed ``hard bugs'', developers can only rely on other features and metadata when debugging.
  
\begin{table}[]
\begin{small}
 \caption{Sample metadata collected by crash error logs}
  \label{tab:metadata}
\begin{tabular}{|p{1.5cm}|p{4.5cm}|p{1.5cm}|}
\hline
 \textbf{Attribute} & \textbf{Explanation} & \textbf{Type} \\ \hline
 Build ID & Build number of the crashing app & Categorical \\ \hline
 OS Version & Version of the mobile operating system & Categorical \\ \hline
 Fd count & Number of open file descriptors & Discrete \\ \hline
 Country & Country associated with the mobile device & Categorical \\ \hline
 Process uptime & Time since app process started & Continuous \\ \hline
 Nav logs & Event navigation sequences before crash & Sequential \\ \hline
 Bi-grams in nav logs & TF-IDF vectorization of bi-grams in nav logs & Continuous \\ \hline
\end{tabular}
\end{small}
\end{table}

\begin{figure}
\caption{Contrived example of a navigation log and its continuous bi-gram features}
\label{fig:example_navlog}
\begin{tabular}{c}
\begin{tabular}{|l|}
\hline
$\textsf{Feed} \xrightarrow{} \textsf{Photos} \xrightarrow{} \textsf{Fundraiser} \xrightarrow{} \textsf{Feed} \xrightarrow{} \textsf{Photos} \xrightarrow{} \textsf{Friends}$ \\
\hline
\end{tabular}
\\ \\
\begin{tabular}{|c|c|}
\hline
\textbf{Bi-gram} & \textbf{TF-IDF weight} \\
\hline
$\textsf{Feed} \xrightarrow{} \textsf{Photos}$ & $2 * 0.01$ \\
$\textsf{Photos} \xrightarrow{} \textsf{Fundraiser}$ & $1 * 9.85$ \\
$\textsf{Fundraiser} \xrightarrow{} \textsf{Feed}$ & $1 * 7.42$ \\
$\textsf{Photos} \xrightarrow{} \textsf{Friends}$ & $1 * 1.25$ \\
$\textsf{Video} \xrightarrow{} \textsf{Feed}$ & $0 * 0.05$ \\
$\ldots$ & $0$ \\
\hline
\end{tabular}
\end{tabular}
\end{figure}

Table~\ref{tab:metadata} shows a subset of the device metadata features that are logged in crash reports.
One feature that developers find particularly useful for dealing with hard bugs are {\em navigation logs}.
A navigation log is a sequence of app surfaces that a user interacted with prior to experiencing the crash.
Figure~\ref{fig:example_navlog} shows an example of a navigation log where the user transitioned from the {\sf Feed} surface to {\sf Friends}. Bi-grams extracted from a navigation log show the source and destination surface of a single navigation event.
These bi-grams help localize crash insights to certain parts of the whole sequence and help reasoning about individual navigation events.

For instance, certain navigation events (bi-grams) tend to occur more commonly than others -- say, navigating to or from {\sf Feed} is more common than {\sf Fundraiser}.
Given a navigation log, this information can be quantified by using the TF-IDF weight of the bi-grams in the log.
TF-IDF~\cite{Robertson:04} is a well-known method in information retrieval to filter the more important features of textual documents from noise.
The TF-IDF weight of a bi-gram denotes how often it appears in the entire corpus of navigation events as opposed to a particular log. Figure~\ref{fig:example_navlog} illustrates
an example.
A high weight for a bi-gram indicates that it is more important to this navigation log, i.e., less common in the entire corpus.
Thus, each bi-gram in the corpus can be considered a feature that can take on any positive real value for each navigation log.

While many features in Table~\ref{tab:metadata}, such as country, are categorical (i.e., their values come from a finite set), TF-IDF encoded bi-gram features are continuous. Categorical features are quite amenable to CSM, whereas continuous features pose several challenges to traditional CSM.

\subsubsection{The Continuous Problem}

The original STUCCO algorithm~\cite{Bay:1999} for CSM generates contrast sets based on {\em categorical} features, such as user locale or CPU model, and Castelluccio \etal ~\cite{Castelluccio:2017} applied it on such features.
In many real world settings, however, crash reports include both categorical variables such as country, and numerical variables (discrete or continuous) such as file descriptor counts and memory usage.

Most applications of CSM to continuous features rely on entropy-based discretization methods, i.e., splitting the continuous domain into discrete intervals and treating them as categorical values.
Following the seminal STUCCO algorithm, Bay proposed an initial data discretization method for CSM~\cite{Bay:2000}.
Simeon and Hilderman proposed a slightly modified equal width binning interval to discretize continuous variables~\cite{Simeon:2008}.

In practice, however, discretization of continuous features has several drawbacks.
First, it greatly increases the number of candidate contrast sets.
Each discretized bin results in a new candidate contrast set, and computation can become prohibitively expensive with a large number of continuous features. Section ~\ref{sec:results} 
presents empirical results which 
quantify this computational cost.

\begin{figure}[h]
  \centering
  \includegraphics[width=0.8\linewidth]{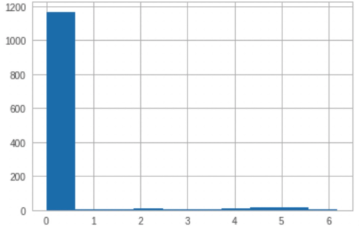}
  \caption{Frequency Distribution of TF-IDF encodings of event sequences}
  \Description{BM25 histogram}
  \label{fig:histogram}
\end{figure}

Second, discretizing continuous data may yield results that are difficult to interpret. Figure ~\ref{fig:histogram} shows the histogram of TF-IDF scores of navigation event sequences. Let's consider the case that we use equal-width bins of width 0.5, and find an arbitrary set of bins, say (0, 0.5), (2, 2.5), and (5, 5.5), are statistically significant, but not other intervals. Developers may not find these results actionable, as the results may reflect the choice of cut-off points more than underlying patterns in the data.

Finally, any form of discretization leads to information loss, especially at the tails. In figure ~\ref{fig:histogram}, the strong skew in distribution with a large proportion of zero values may drive discretization, with the second smaller hump at (4.5, 5.5) going unnoticed. In this case, most discretized contrast sets cannot represent the magnitude of mean differences between two groups, even if mean difference provides important debugging information to developers. In our context, if repeated navigation events that are rare overall lead to out-of-memory crashes, mean difference in number of navigation events would convey important debugging context without information loss.

These drawbacks of discretization provide motivation for developing a continuous version of contrast set mining.

\section{Continuous Contrast Set Mining}
\label{sec:algorithms}
Addressing the limitations just described, we propose the CCSM algorithm, which
applies CSM directly to continuous data.
Additionally, we define separate and unified anomaly scores for continuous and categorical contrast sets. Recognizing that real world datasets are frequently mixed, our \textit{unified anomaly score} is the first ranking algorithm that produces a normalized comparison of the two anomaly definitions.
Finally, we describe confidence intervals on contrast sets, and how we translate them into interpretable findings.

\subsection{Base CCSM Algorithm}
The CCSM algorithm adopts the same structure
as STUCCO described earlier in Section~\ref{sec:overview}, with modifications
to reason about sets of continuous attributes. 
As previously discussed, STUCCO requires
discretization to handle continuous or discrete features with numerical ranges. The CCSM algorithm instead reasons directly
about continuous contrast sets without
introducing discretized bins.

For CCSM, the input consists of a set of $k$-dimensional numerical vectors, where $k$ is the number of continuous variables, partitioned into mutually exclusive groups (\MID{}s).  A contrast set is either a single continuous variable or a set of continuous variables.  We start by considering single continuous variables.

As in the original STUCCO algorithm, we consider a contrast set a deviation if it is both \emph{significant} and \emph{large}. We develop counterparts for these two conditions in the continuous domain.

We define a contrast set to be \textit{significant} or statistically significant if it passes a \textit{one-way ANOVA F-test}.  The null hypothesis of the one-way ANOVA F-test is that the average value of the contrast set is the same across all the groups.  Then, we reject the null hypothesis if at least two groups have different average values for this contrast set.  This is a natural counterpart to the \textit{Pearson\textquotesingle s chi-squared test} used in the original STUCCO algorithm, which tests the null hypothesis that the percentage of the contrast set is the same across all the groups. As with the STUCCO algorithm, we apply a set of pruning heuristics and apply the Bonferroni correction to reduce the likelihood of type 1 errors in tree-based search and testing.

We define a contrast set to be \textit{large} or practically significant if there exist two groups such that the difference of the average values of the contrast set in these two groups is greater than some user-defined threshold $\delta$.  This definition also mirrors that in the original STUCCO algorithm, where a contrast set is defined to be large if the percentage difference of a contrast set in two groups is larger than some threshold.

Given these definitions of \textit{significant}
and \textit{large}, we apply the STUCCO tree search
algorithm to efficiently search for conjunctions
of contrast sets that distinguish a particular
group of vectors (a \MID{}) from the rest of
the population.   Algorithm~\ref{euclid}
shows pseudo-code for the CCSM algorithm,
with the base algorithm starting on line 9.
 Algorithm~\ref{euclid} additionally includes
 some details specific to mining navigation
 logs -- subsequent sections discuss these
 details.

\makeatletter
\def\BState{\State\hskip-\ALG@thistlm}
\makeatother

\begin{algorithm}
\caption{Continuous Contrast Set Mining on Navigation Sequences}\label{euclid}
\begin{algorithmic}[1]
\Procedure{CCSM algorithm}{}
\State $Q \gets \textit{initial candidate set of n-grams in S}$
\State Result set $R \gets \emptyset$
\While{\textit{Q is not empty }}
\BState \emph{preprocess}:
\For{\textit{each q in Q}}
\State $IDF(q) = ln( \frac{docCount - f(q)+0.5}{f(q) + 0.5}) $
\State count $f(q, d) $ for each FAD $ d \in D$
\EndFor
\BState \emph{CCSM}:
\For{\textit{each q in Q}}
\State $\texttt{is\_significant} \gets ANOVA(q, X)$
\State $\texttt{means\_difference} \gets \max_{i,j}|mean(q | G_i) - mean(q | G_j)| $
\If {$prune(q)$ is True}
\State continue
\EndIf
\If {$\mathit{is\_significant} \wedge \mathit{means\_difference} > \delta $}
\State append $q$ to $R$
\EndIf
\EndFor
\State $Q \gets \texttt{gen\_candidates}(R)$
\EndWhile

\State \Return $R$
\EndProcedure
\end{algorithmic}
\end{algorithm}

\subsection{Ranking Contrast Sets}
We use Cohen\textquotesingle s \textit{d} as the anomaly score for continuous contrast sets. Cohen\textquotesingle s \textit{d} is a measure of effect size, or more specifically, a measure of the difference between two group means.  The formula is given by $$ d = \frac{\Bar{x}_1-\Bar{x}_2}{s}$$ where $\Bar{x}_1$ and $\Bar{x_2}$ are the two sample means, and $s$ is the pooled standard deviation, defined as 
$$ s = \sqrt{\frac{(n_1-1)s_1^2 + (n_2 - 1)s_2^2}{n_1 + n_2 - 2}} $$
where $s_1$ and $s_2$ are the two sample standard deviations.  The anomaly score can be seen as the standardized mean difference on contrast set X between group A and all the other groups.  If we see $d=1$, we know that the two means differ by one standard deviation; $d=0.5$ tells us that the two means differ by half a standard deviation, and so on. 

Table~\ref{tab:freq} shows how we convert measures of effect size into descriptive interpretations that can be displayed in crash analysis UI. The higher the anomaly score is (which means X has larger values inside A than outside A), the more prominent the contrast set X is to group A.

\begin{table}
  \caption{Qualitative evaluations of Effect Size}
  \label{tab:freq}
  \begin{tabular}{ccl}
    \toprule
    Magnitude&Effect Size (d)\\
    \midrule
    very small& 0.01\\
    small & 0.2\\
    medium & 0.5\\
    large & 0.8\\
    very large & 1.2\\
    huge & 2\\
  \bottomrule
\end{tabular}
\end{table}

For every group, we score each CCSM contrast set by its anomaly score.  We then submit the top deviations for each group, ranked by their anomaly scores. Note here that we can extend the use case to surface contrast sets where group A has small values rather than large values compared to other groups by ranking the contrast sets by the absolute value of the anomaly scores instead.

\subsubsection{Unifying Continuous and Categorical Contrast Sets}

As described in Section ~\ref{sec:overview}, STUCCO relies on categorical inputs. Inspired by STUCCO's tree search strategy, CCSM defines a novel discriminative pattern mining algorithm on continuous datasets. However, it is more likely that real world datasets contain mixed data types; Section ~\ref{sec:csm_industry} explains that crash data is high dimensional and heterogeneous. To enable comparisons between continuous and categorical contrast sets, we need a unified ranking algorithm that uses both types of features.  The key is to make the anomaly scores for both types comparable.

To do so, we propose a new definition of anomaly score for categorical contrast sets. For a categorical contrast set X and group A, we want to see if the percentage (or support) of X is higher inside A than outside A.  To achieve this, we use Cohen\textquotesingle s \textit{h } as the anomaly score, which is the difference between two proportions after an \texttt{arcsine} root transformation.  Specifically, its formula is given by
$$ h = 2(arcsin\sqrt{p_1} - arcsin\sqrt{p_2}) $$
where $p_1$ and $p_2$ are the two sample proportions.  

The goal of employing this transformation is to mitigate bias, where very rare contrast sets are disproportionately surfaced as significant and large.  Without any transformation, the variance of the proportion is given by $p(1-p)$; the variance is small when the proportion is close to 0.5 and large when it is close to 0 or 1.  For example, if two proportions are both around 0.5, it is easy to detect their difference; if they are both close to 0, it is hard to detect their difference.  The \texttt{arcsine} root transformation stabilizes the variance for all proportions, and hence, makes all proportion differences equally detectable.  Cohen\textquotesingle s \textit{d} can be seen as standardized difference of means, while Cohen\textquotesingle s \textit{h} can be seen as difference of standardized proportions.

Note that Cohen\textquotesingle s \textit{h} has the same rule of thumb for categorizing the magnitude of the effect size as Cohen\textquotesingle s \textit{d}, making the two comparable to each other.

\subsubsection{Comparing anomaly score to percent difference in supports}

The percent increase for a categorical contrast set to a group is defined as
$$ \frac{\text{observed support} - \text{expected support}}{\text{expected support}} $$

Here, observed support is defined as the percentage of a contrast set in a certain group, whereas expected support is defined as the percentage of a contrast set across all the groups.  By Taylor\textquotesingle s Theorem, the new anomaly score (Cohen\textquotesingle s \textit{h}) for categorical contrast sets can be approximated by

$$ \frac{\text{observed support} - \text{expected support}}{\sqrt{\text{expected support} (1 - \text{expected support})}} $$
up to some constant factor, under certain conditions.  The major difference between these two definitions lies in the denominator, or how we standardize the difference between observed support and expected support.
The new anomaly score not only comes with nicer statistical properties, but also mitigates the bias in ranking contrast sets using the old anomaly score.  When a contrast set is very rare (i.e., expected support is very close to zero), the anomaly score of this contrast set, under the previous definition, will universally tend to be high.  This makes ranking unfair to those comparatively frequent contrast sets (those with expected support not so close to zero).

However, under the new definition, when expected support is very close to zero, square root of expected support will cause it to deviate zero while $(1 - \text{expected support})$ stays very close to one.  Thus it gives a larger denominator and the difference in the numerator is not over-standardized.

Moreover, the new anomaly score is closely connected to the Pearson\textquotesingle s Chi Squared test for independence. Specifically, the $\chi^2$ statistic is proportional to
$$ \frac{(\text{observed support} - \text{expected support})^2}{\text{expected support}(1 - \text{expected support})}$$
which is simply the square of the approximation of Cohen\textquotesingle s \textit{h} displayed above.  However, instead of using the Chi Squared test, we believe Cohen\textquotesingle s \textit{h} better suits our needs. The reason is that the anomaly score based on Cohen\textquotesingle s \textit{h} is positive only if observed support is larger than expected support, while the $\chi^2$ statistic is always positive, even when observed support is smaller than expected support. While these "negative contrast sets" can still be interesting~\cite{Wong:2005}, recall that in our context we want to select features which are more frequent inside a group than outside this group.

\subsection{Confidence Intervals}

For each anomaly we find, we provide confidence intervals for the mean and percentage difference to increase actionability of results. 

The first confidence interval is provided for the effect size (Cohen\textquotesingle s \textit{d} or Cohen\textquotesingle s \textit{h}).  For example, a Cohen\textquotesingle s \textit{d} of 0.5 tells us that based on the sample, there is a medium difference between current group and rest of the groups on a certain feature.  If we further obtain a 95\% confidence interval of (0.48, 0.52), we know that the sample is a good representation of the population, and we can be quite confident that there is a medium difference.  However, if the interval is (0.19, 0.81), we are not certain that the difference is of a medium size; it could actually be large or small.

We use standard methods for constructing confidence intervals for Cohen\textquotesingle s \textit{d} and Cohen\textquotesingle s \textit{h}~\cite{Cohen:92}.  We apply Bonferroni correction on up to 20 anomalies for each group. 

Another confidence interval is provided for the mean and percentage differences. We refer to (observed mean - expected mean) for continuous contrast sets and (observed percentage - expected percentage) for categorical contrast sets. This gives a better idea of what the difference looks like at the original scale without any standardization. For example, if a given feature has an expected support of 10\% and an observed support of 50\% in certain group, the difference of these two percentages is simply 40\%; a confidence interval of (39\%, 41\%) would reassure developers that the percentage difference is estimated relatively precisely, compared to a confidence interval of (25\%, 55\%). 

For continuous contrast sets, we use Welch\textquotesingle s t-interval; for categorical contrast sets, we use Wilson score interval (without continuity correction). As before, we apply Bonferroni correction.

\section{Contrast Set Mining in Practice}
\label{sec:csm_industry}

In this section, we give a broader picture of how CSM would be used in an industrial organization for crash diagnosis.
Particularly, we describe the architecture of mobile app reliability tools at Facebook and how results from CSM can help developers.

\subsection{Mobile App Reliability at Facebook}
\label{subsec:fb_reliability}

\begin{figure*}
  \centering
  \includegraphics[width=0.85\linewidth]{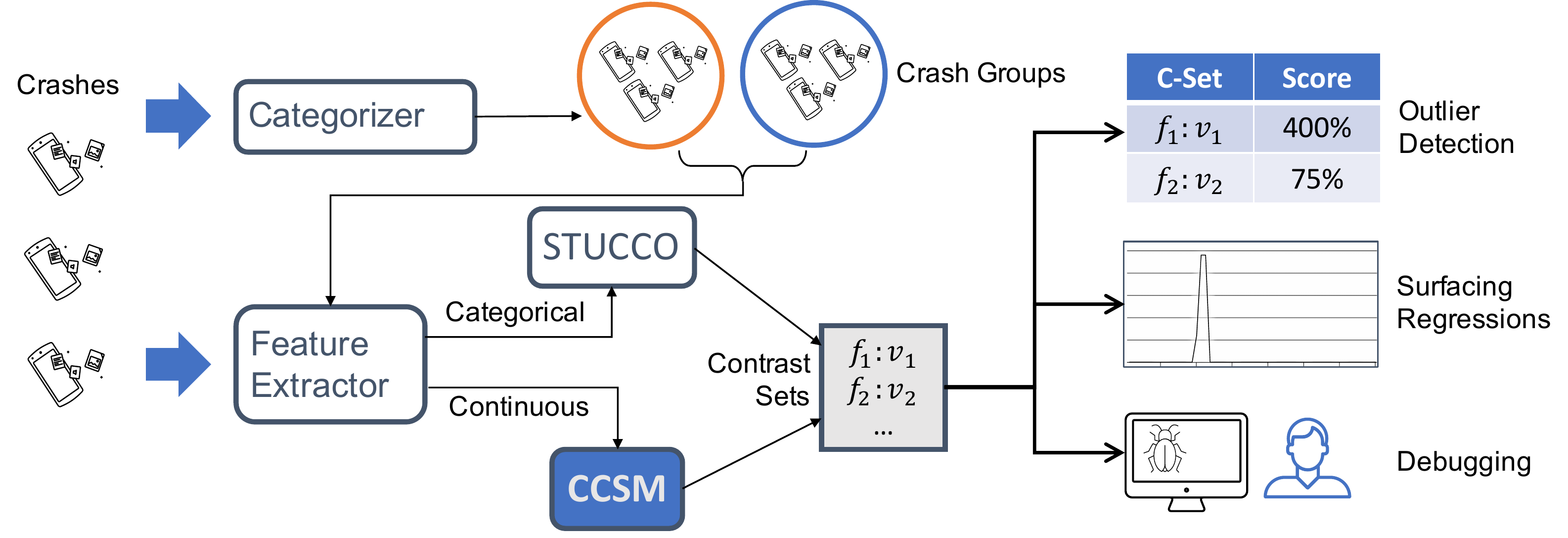}
  \caption{Overview of the crash analysis system}
  \label{fig:system}
  \Description{triage}
\end{figure*}

Figure~\ref{fig:system} shows an overview of Facebook's mobile crash analysis architecture.
When a client experiences a crash, the generated crash report is received by a ``categorizer''.
The job of the categorizer is to assign the crash to a particular group of crashes, such that all crashes in the a group arise from the same root cause bug.
The categorizer makes use of a mix of heuristics and ML-based approaches to compute these groups, and in the end produces a unique signature representing the group to which the crash was assigned.
We will denote this group identifier as \MID{} and use it to refer to a group of crashes.

The crash report is also fed through a feature extractor that extracts metadata features such as the ones in Table~\ref{tab:metadata}.
The features can then be processed by CSM algorithms -- categorical features can be processed by the traditional STUCCO algorithm, and continuous features can be used as input to the CCSM algorithm presented in this paper.
The output of the CSM algorithms are contrast sets, which are conjunctions of feature-value pairs.

Contrast sets are useful for a variety of downstream purposes.
First, prominent contrast sets (based on our notion of anomaly score presented here) are useful for describing groups.
Each \MID{} indexes into an internal issue tracker system where developers monitor spikes of crashes in the \MID{}, create tasks to work on bug fixes and mitigate issues.
Contrast sets are ranked by their anomaly scores, which denotes the degree to which CCSM believes the contrast set to be prominent for the \MID{}, and displayed in the issue tracker UI. Since string keys assigned by categorizers do not have semantic meaning, contrast sets provide interpretable descriptions of groups of crash reports.

Second, they can surface spikes in \MID{}s that would otherwise go under the radar.
For instance, users from a country Y using a particular build version X can be experiencing a spike of crashes.
However, if the actual {\em number} of such crashes is small compared to the size of the \MID{}, the regression is unlikely to be surfaced at the \MID{} level.
CSM would be able to produce the contrast set $\{{\sf country:Y}, {\sf build\_version:X}\}$, filtering on which would reveal the spike.

Finally, contrast sets can provide useful hints to developers debugging crashes in a \MID{}.
Currently, the issue tracker UI displays a simple count of features, which, as we discussed in section ~\ref{sec:intro}, can be misleading.
Contrast sets, on the other hand, surface features that statistically distinguish a \MID{} from others, guiding developers to which features are more likely to be related to the \MID{}.

\subsection{Usability of Contrast Sets}

As noted by Webb~\etal~\cite{Webb:2003}, visualizing contrast sets is a challenging open problem in the pattern mining space. We identified two key pain points to the consumption of contrast sets:
\begin{itemize}
    \item \textit{Actionability}: How much can we trust these findings? How likely are developers to act upon this information? Developers value transparency in ML models and want to quantify uncertainty. To address this, we provide a notion of \textit{confidence} in Section ~\ref{sec:algorithms} so that end users can assess the representativeness of our findings.
    \item \textit{Interpretability}: Anomalous features should be immediately obvious to developers inspecting a group of crashes. This becomes a more challenging task when we consider conjunctive contrast sets as well.
\end{itemize}

To improve interpretability of statistical measures, we propose an alternative to the current visualization for crash statistics that only highlights meaningful statistical differences. As noted in Section 3.1, simple frequency counts can be misleading.

\begin{figure}
\begin{tabular}{c}
  \includegraphics[width=0.8\linewidth]{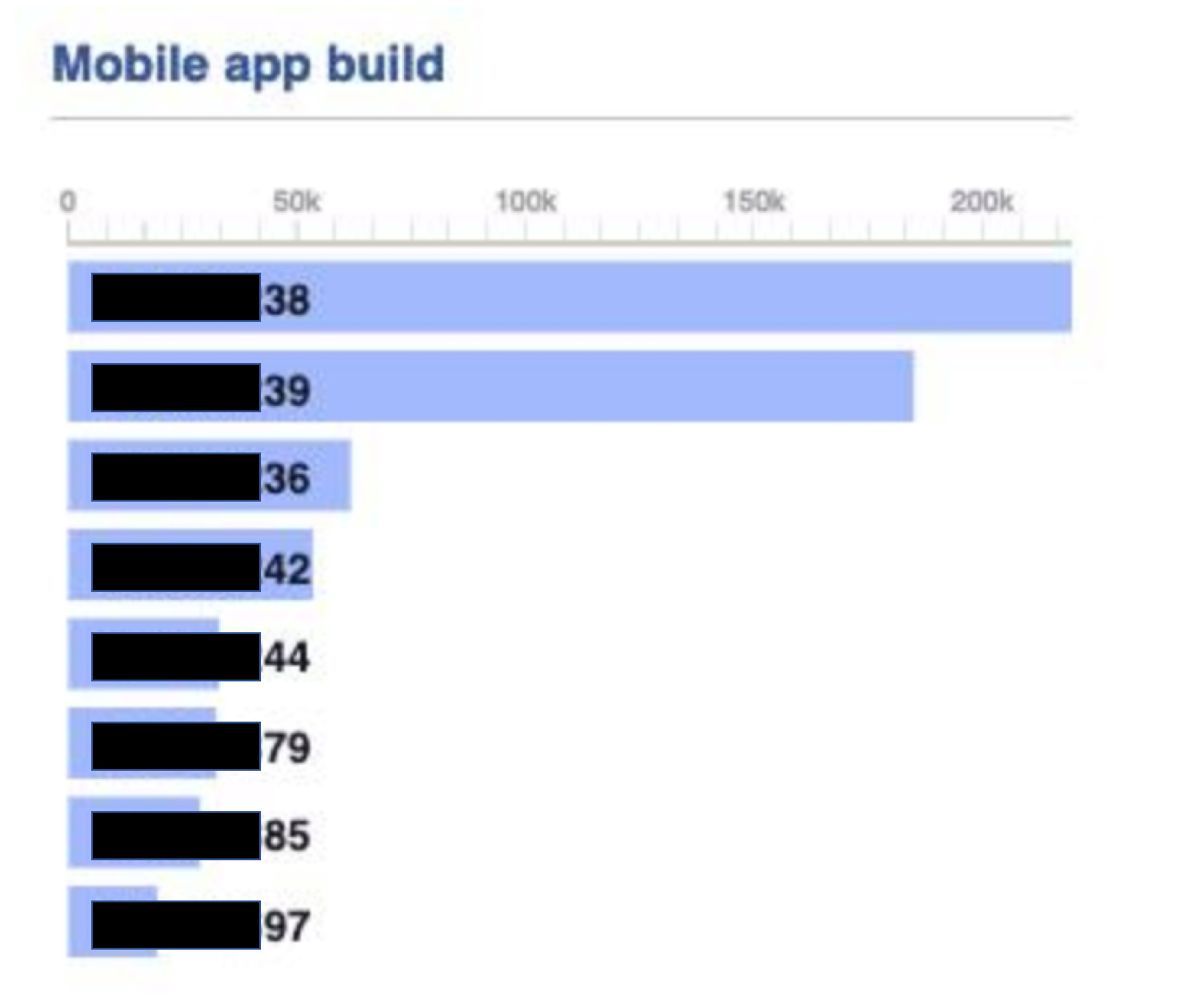} \\ (a) \\ \\
  \begin{tabular}{ccc}
  \hline
  {\bf Feature} & {\bf Percent Deviation} \\
  \hline
  \{{\sf app\_build: 44}\} & $150\%$ \\
  \{{\sf os\_version: 4.0, country: US}\} & $102\%$ \\
  \{{\sf connection\_type: WIFI}\} & $45\%$ \\
  \{{\sf device\_brand: Samsung}\} & $38\%$ \\
  \ldots \\ \hline
  \end{tabular}
\end{tabular} \\ (b) \\
\caption{(a) Counts of app builds for a particular \MID{}, (b) Anomalous features found by CSM}
\label{fig:app-builds}
\end{figure}

To illustrate the consequences of misleading visualizations, let us go through a realistic example.
Figure ~\ref{fig:app-builds}(a) shows a ranking of mobile app builds for a particular \MID{} by \textit{count}.
Simply judging by the prevalence of build 38 in the \MID{}, one might incorrectly conclude that it is closely associated with the bug.
CSM, on the other hand, revealed that build 38 is {\em expected} to be prevalent because it also occurred with similar frequencies in other \MID{}s, perhaps being the most used build of the app.
Instead, it found that build 44 is the most anomalous as its anomaly score is 150\% above expected thresholds, as shown in Figure~\ref{fig:app-builds}(b).
Engineers working on the \MID{} validated this insight, and eventually fixed the bug by gating out this build.
This is an example of how CSM results can be presented to developers to aid them in debugging.
In addition to UI tooling, we can integrate CSM results into the debugging workflow through scripts posting daily findings to oncall groups, and bots that automatically comment on open tasks associated with \MID{}s. 

\section{Evaluation}
\label{sec:results}

To evaluate the proposed techniques, we consider the following questions:
\begin{itemize}
\item \textbf{RQ1}: Does CCSM have lower computational cost than existing approaches? Is CCSM efficient enough to scale to high cardinality, high dimensional datasets?
\item \textbf{RQ2}:  Does contrast set mining help diagnose crash reports in our environment?
\item \textbf{RQ3}:  Does our definition of anomaly score add value over previous ranking techniques relying on percentage differences?
\end{itemize}

\subsection{Implementation Details}
To evaluate model efficiency, we collected on the order of $60k$ field crashes each day from the week of September 10, 2019 to September 16, 2019. The data consists of iOS Out-of-Memory (OOM) crashes from the core Facebook mobile app. As discussed in Section ~\ref{sec:overview}, OOM crashes are difficult bugs without accompanying stack traces, and are thus good candidates for contrast set mining. For each crash, we have metadata such as the device OS and app build as in Table ~\ref{tab:metadata}, which include categorical, discrete, and continuous data types. Many crash reports include a sequence of navigation events, such as the example shown in Figure ~\ref{fig:example_navlog}. We follow the procedure described in Section ~\ref{sec:overview} to extract bigrams and embed them using TF-IDF weights, generating thousands of continuous columns for each dataset.

We simulate using CCSM at different points in time in a production setting. For our baseline, we use a standard implementation of STUCCO with equi-width binning. We compare CCSM to the binning approach with two different number of bins, 3 and 10. We demonstrate that discretization of bigrams is much slower than directly applying continuous contrast mining and generates lower quality contrast sets.

To mitigate the effects of uneven group sizes on our evaluation, we use stratified sampling to control the number of crashes fetched for each \MID{} in expectation. For each day, we ran our evaluation on $1k$, $10k$ and $60k$ crashes. At the end of each run, we recorded the runtime and inspected the output contrast sets to ensure quality of our results. We set an upper bound of $3600s$ (1 hour) for execution time; runs that exceeded this threshold were terminated due to high memory usage.

\subsection{RQ1: Analysis of Execution Times}

\begin{figure*}
	\begin{minipage}{.45\linewidth}
		\includegraphics[width=\linewidth]{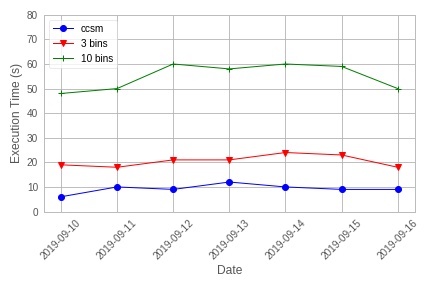}
		\begin{center}
			(a) $1k$ crash reports.
		\end{center}
	\end{minipage}
	\hspace{1em}
	\begin{minipage}{.45\linewidth}
		\includegraphics[width=\linewidth]{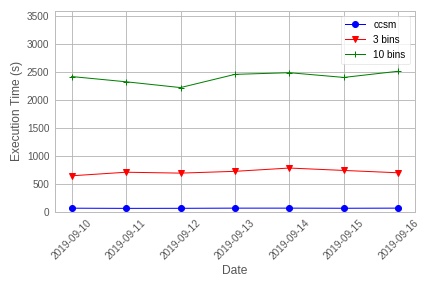}
		\begin{center}
			(b) $10k$ crash reports.
		\end{center}
	\end{minipage}

	\vspace{1em}
	\begin{minipage}{.45\linewidth}
		\includegraphics[width=\linewidth]{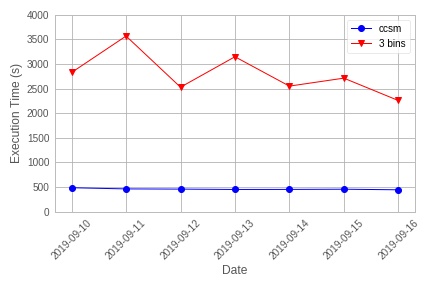}
		\begin{center}
			(c) $60k$ crash reports.
		\end{center}
	\end{minipage}
	\hspace{1em}
	\begin{minipage}{.45\linewidth}
		\includegraphics[width=\linewidth]{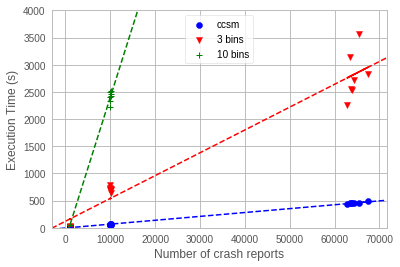}
		\begin{center}
			(a) Run time vs. Input size.
		\end{center}
	\end{minipage}

	\caption{Execution times for different contrast mining implementations. The blue line is the CCSM approach.}
	\label{fig:exec-times}
\end{figure*}

Figure ~\ref{fig:exec-times} shows the execution times of different CSM models over time. The results vary over time because at each time interval, we collect a new set of crash reports. Since stratified sampling controls sample sizes in expectation, the cardinality of our dataset varies slightly as well across runs.

Both baseline approaches perform poorly when compared to CCSM. We find that discretization suffers from acute scalability problems. This is especially true of smaller bin widths; discretization with 10 bins consistently exceeded the time limit for the $N=60k$ dataset. On average, CCSM achieves a \textbf{40$\mathbf{x}$ speedup} over discretization with 10 bins and a $10x$ speedup over discretization with 3 bins. Prior work observed that using fewer bins generally leads to faster execution~\cite{Zhu:2015}, but outputs fewer contrast sets and incurs greater information loss from bucketing. We validate these findings empirically in our results. 

It should be noted that in pattern mining research, it is common to partition data ranges into hundreds of bins~\cite{Zhu:2015}. Since finer partitions would only further increase computational costs of discretization, we demonstrate our efficiency improvements on relatively simple baseline approaches.

\subsection{RQ2: Validation of results}

We considered 24 high priority crash tasks (all closed) generated from July-September 2019, where contrast mining generated findings. 16 of the tasks involved hard bugs, where stack traces were unavailable or difficult to parse. For these tasks, we selected 32 contrast sets we generated with the highest ranked anomaly scores. We have manually analyzed this set of crashes and the discussion and code changes that are attached to them, along with contrast mining findings. We labelled each contrast set as directly useful, relevant or compatible, and not helpful. We found 12 cases where the tool surfaced interesting patterns that were directly useful to the crash resolution; 18 cases where the tool generated compatible results but were not sufficient to root cause the bug, and 2 cases where our mining tool was not helpful. 

\subsection{RQ3: Improvement in Usability of Anomaly Scores}

Past work on pattern mining in a classification context has the goal of maximizing predictive accuracy, such as discretizing continuous attributes for a naive Bayesian classifier ~\cite{Bay:2000}. In pattern mining we analyze the data in an exploratory fashion, where the emphasis is not on predictive accuracy but rather on finding previously unknown patterns in the data.

We thus use an example to illustrate how the new anomaly score and the current practice of using percent difference between expected and observed supports rank the features differently. We ran contrast set mining using data from July 2019. Figure 7 contains anomalies found for a regressing \MID{}, and the anomalies are ranked by the new anomaly score and the original percent difference based ranking.

It is easy to see that the original anomaly score is in favor of finding anomalies with low expected support, such as \texttt{time\_since\_init\_ms}: (0, 150000) and \texttt{background\_time\_since\_init\_ms}: (0, 1000).

The new anomaly score gives a higher rank to anomalies where the expected support is not extremely low, such as \texttt{major\_app\_version: 229}. This mitigates bias towards rare feature sets.

\begin{figure*}
\caption{Comparison of statistical ranking definitions}
  \begin{tabular}{c|c|c|c|c}
    \toprule
    Features& Expected Support & Observed Support & Weighted Anomaly Score & Percent Difference\\
    \midrule
    app version = 2 & 0.68 & 1 & 1.364 & 0.472 \\
    app build = 123 & 0.68 & 1 & 1.363 & 0.471 \\
    time since init = (0, 150000) & 0.081 & 0.214 & 0.518 & 1.657\\
    OS version = 12 & 0.742 & 0.858 & 0.355 & 0.156 \\
    background time since init = (0, 1000) & 0.086 & 0.172 & 0.341 & 1.008 \\
  \bottomrule
\end{tabular}
\end{figure*}

\subsection{Early Experiences}
Since experimenting with contrast set mining on field crashes, we have found numerous cases where CCSM surfaced important insights on hard issues, and in some cases, found the root cause of groups of crashes. We find that the analysis of embedded navigation event sequences using CCSM adds significant value to crash analysis using STUCCO. By pinpointing specific frames, CCSM is able to provide more actionable insights than analysis on categorical variables alone. We describe several instances where both CCSM and STUCCO helped guide the debugging process below.

\begin{itemize}
\item \textit{Root Cause for Hard Bugs}. Testing CCSM on production data over multiple days when a group of crashes was the most prevalent among users, we found that navigation to and from a navigation module related to comments showed up within the top five anomalous features consistently. Product engineers confirmed that the fix for the issue involved the navigation events surfaced by CCSM. This crash is an example of OOM errors, which are especially difficult to debug due to the lack of stack traces (see Section ~\ref{sec:overview}). This is a case where continuous CSM makes it possible to analyze hard bugs due to its scalability to high dimensional datasets.

    \item \textit{Issue Discovery}. Contrast Mining detected that a certain app build was highly anomalous and experiencing high crash rates. For the corresponding \MID{}, contrast mining data showed that the number of app crashes we observed with this build was 56\% higher than expected. This build number represents the x86 build type. This is an example of an underlying issue that otherwise can be left unnoticed because the cohort of affected devices is very small, and signal was diluted because the build type failures are spread across 5-6 different \MID{}s.
    \item \textit{Describing Crash Groups}. Contrast mining finds statistically significant deviations to help pinpoint the root cause of crashes. We ran contrast mining on a \MID{} associated with a high priority task. We then contacted the task owner with the list of anomalous features. The mobile engineer found the information very helpful as a way to differentiate between normal behavior and statistically significant differences. Specifically, one highly anomalous contrast set indicated that connection class was poor. ``The fix that I proposed is based on the assumption that the network is slow, and this confirms that.''
\end{itemize}

\section{Threats to Validity}
\label{sec:discussion}

Our application of CCSM focused on a limited set of continuous variables, such as TF-IDF from navigation events and the active time of the mobile device, where mean differences are useful statistics to focus on. The usability of the algorithm and the usefulness of the results depend heavily on the context, specifically on whether binned continuous variables or overall mean of the variables provide more useful information.   

The qualitative case studies have been performed by authors to evaluate the algorithm in production setting with engineers and may suffer from selection bias as positive results are more likely to be reported.

\section{Related Work}
\label{sec:related}

There is a growing body of research applying machine learning techniques to crash triaging and resolution. Information retrieval based bug localization techniques extract semantic information from crash stacks, and have been shown to scale to large project sources with low cost text analysis ~\cite{Rao:2011}. Wu~\etal located crash-inducing changes by training classification models on candidates extracted from buckets of crash reports~\cite{Wu:2018}.

The problem of bucketing crash reports has been well studied in literature. Dhaliwal~\etal~\cite{Dhaliwal:2011} found that crash clusters containing reports triggered by multiple bugs took longer to fix, and proposed grouping crashes using the Levenshtein distance between stack traces. Campbell~\etal~\cite{Campbell:2016} found that off-the-shelf information~retrieval techniques outperformed crash deduplication algorithms in categorizing crashes at scale.

For the most part, the above approaches focus on mining information at the trace level, or individual crash reports. Our approach focuses on analyzing characteristics of groups of crashes in aggregate to aid developers in crash triaging and resolution. Castelluccio~\etal~\cite{Castelluccio:2017} presented the first application of CSM to the problem of crash group analysis. To the best of our knowledge, we are the first to bring contrast set mining to the continuous domain.

\section{Conclusion}
\label{sec:conclusion}
App crashes are severe symptoms of software errors, causing significant pain for both end users and oncall engineers. Maintaining app health is therefore one of the top priorities of large software organizations. We propose CCSM, a novel pattern mining algorithm for continuous data that scales to datasets with thousands of features. We found that automated crash analysis can detect anomalous patterns that are difficult to identify with manual inspection, and that the analysis of continuous features greatly adds value to existing categorical data mining approaches. 
\subsection{Future Work}

\subsubsection{Correlations between features}
For features that frequently co-occur independent of group membership, it is helpful to surface this information to developers, or down-weight the anomaly scores of these contrast sets. We are working on developing an algorithm to automatically detect feature dependencies using cross feature entropy.

\subsubsection{Parametric Assumptions}
Both the \textit{one-way ANOVA} and \textit{two-sample t-test}, to which Cohen’s d is connected to, are known to be robust to the normality assumption. However, \textit{one-way ANOVA} also assumes homogeneity of variances among the groups, which may not be realistic for distributions of field crashes. When this assumption is violated, we can use non-parametric methods such as a \textit{Kruskal-Wallis H Test} to test statistical significance instead.

\subsubsection{Pruning rules}

We rely heavily on heuristics and domain expertise to set user-defined thresholds to prune nodes that are not \emph{large}. Understanding the sensitivity of the performance of contrast learning to these thresholds and exploring heterogeneous thresholds for different sets of features based on domain expertise will be useful contribution to the literature.

\bibliographystyle{ACM-Reference-Format}
\bibliography{csm}

\end{document}